\documentclass[sigconf,screen,nonacm,review=false,timestamp=false]{acmart}

\usepackage{subcaption}
\usepackage{amsmath}
\AtBeginDocument{%
  }

\begin{document}

\title{Follow Nudges without Budges: A Field Experiment on Misinformation Followers Didn’t Change Follow Networks}

\author{Laura Kurek}
\email{lkurek@umich.edu}
\affiliation{%
  \institution{University of Michigan}
  \country{USA}
}
\author{Joshua Ashkinaze}
\email{jashkina@umich.edu}
\affiliation{%
  \institution{University of Michigan}
  \country{USA}
}
\author{Ceren Budak}
\email{cbudak@umich.edu}
\affiliation{%
  \institution{University of Michigan}
  \country{USA}
}
\author{Eric Gilbert}
\email{eegg@umich.edu}
\affiliation{%
  \institution{University of Michigan}
  \country{USA}
}

\renewcommand{\shortauthors}{Kurek et al.}
\renewcommand{\shorttitle}{Follow Nudges without Budges}

\begin{abstract}
Can digital ads encourage users exposed to inaccurate information sources to follow accurate ones? We conduct a large-scale field experiment (N=28,582) on X, formerly Twitter, with users who follow accounts that spread health misinformation. Participants were exposed to four ad treatments varied on two dimensions: a neutral message versus a persuasive message appealing to values of independence, and a request to follow a health institution versus a request to follow a health influencer. We term this ad-based, social network intervention a \textit{follow nudge}. The ad with a persuasive message to follow a well-known health institution generated significantly higher click-through rates than all other conditions (Bonferroni-corrected pairwise tests, all p<0.001). Given the overall low click-through rate across treatments and the high cost of digital advertising infrastructure on X, however, we conclude that our proposed intervention---at least in its current ad-based format---is not a cost-effective means to improve information environments online. We discuss challenges faced when conducting large-scale experiments on X following the platform's ownership change and subsequent restrictions on data access for research purposes.
\end{abstract}




\maketitle

\section{Introduction}
Social media networks shape misinformation exposure and consumption, as follower networks influence the default content that users see. Users who follow spreaders of health misinformation will log onto their account and likely encounter more health misinformation in their feeds, due in part to algorithmic and network homophily factors~\cite{cinelli_echo_2021}. But how entrenched are these networks? Prior work has found that people rarely unfollow those they have chosen to follow online, including health misinformation spreaders~\cite{kivran-swaine_impact_2011, ashkinaze_dynamics_2024}. Here we test the inverse: \textbf{Are social media users who follow spreaders of health misinformation open to following higher-quality health sources?} We test this by running a large-scale field experiment (N=28,582) on X (formerly Twitter), targeting followers of known health misinformation spreaders.

\begin{figure*}[h]
\centering
\includegraphics[width=0.65\textwidth,height=0.7\textheight,keepaspectratio]{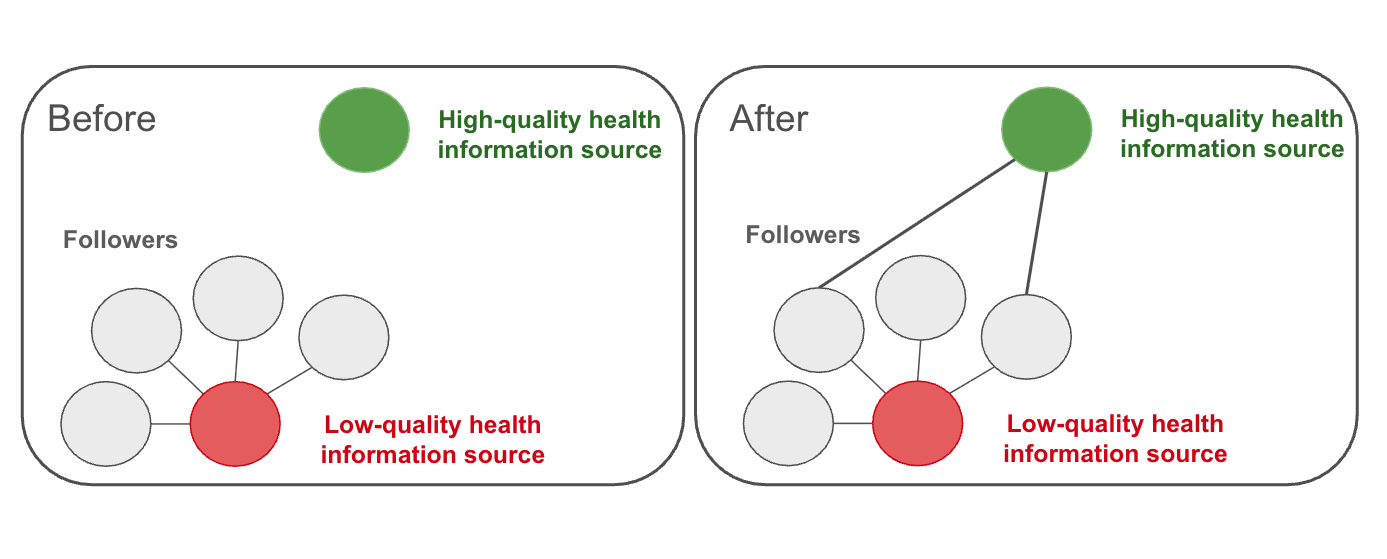}
    \captionsetup{width=0.65\textwidth}
    \caption{The \textit{follow nudge} intervention intends to add high-quality information ties within low-quality information networks.}
    \label{fig:study_design}
\end{figure*}

We propose an ad-based, social network intervention of \textit{follow nudges}, where followers of health misinformation spreaders are prompted to follow an authoritative health source via targeted digital advertisements. See Figure~\ref{fig:study_design} for an illustrative diagram of our proposed network intervention. Participants were exposed to four ad treatments varied along two dimensions: a neutral message versus a persuasive message appealing to values of independence, and a request to follow a U.S. health institution versus a request to follow a health influencer. This intervention aims to diffuse the follower's exposure to a low-quality network tie by adding a high-quality network tie, lowering the search costs of finding high quality information. 

Our work follows scientific recommendations to develop nudges for pro-social purposes~\cite{lorenz-spreen_systematic_2022}, with two important design advantages over prior misinformation interventions: 
\begin{enumerate}
    \item \textit{Persistence}: The operating mechanism---an addition of high-quality network tie---would likely not revert itself, making it persistent by design. Other misinformation interventions, such as prebunking or accuracy nudging, have been shown to have decaying effects~\cite{basol_towards_2021, compton_persuading_2016, hill_how_2013}.
    \item \textit{Additive approach}: Instead of breaking existing social ties, we encourage social media users to diversify their networks by following someone new. Instead of trying to label or remove problematic information---which can come with backfire effects~\cite{lewandowsky_misinformation_2012}---we encourage the addition of high-quality information into the information ecosystem.
\end{enumerate}

Determining whether online social networks can be altered by \textit{follow nudges} is important regardless of the answer. If yes (misinformation followers are receptive to following high-quality information sources), there are potential positive first-order and second-order effects. For the user who decided to follow a health authority, encountering high-quality information may make false information seem less plausible~\citep{bode_see_2018}. For the followers of that user, they might also be exposed to high-quality health information, as algorithmic and network factors propagate content through social connections. If no (misinformation followers are not receptive to calls to follow high-quality sources), then it would suggest that misinformation networks are resistant to ad-based interventions encouraging new social ties. Such resistance to follow new accounts would be compounded by the tendency of users to rarely unfollow those they have already followed, particularly health misinformation spreaders~\cite{kivran-swaine_impact_2011, ashkinaze_dynamics_2024}. Finally, ad-based interventions in general are important to explore in today's data-constrained landscape. Despite recent API restrictions, digital advertising infrastructure remains accessible to researchers, and offers potential opportunities for experiments: targeting capabilities, custom audience lists, and scalability.
 
Our findings suggest the latter `no' scenario: health misinformation networks are resistant to our ad-based \textit{follow nudge} interventions. Despite targeted advertising efforts, misinformation followers infrequently clicked the ads and rarely chose to follow high-quality sources (impressions=28,582, click-through rate=1.04\%, follows $\approx3$). The click-through rate (CTR) of 1.04\% represents an upper bound of potential follows, as users must first click the ad, and then click the `follow' button. Due to major changes to the X API in the midst of this study, we were unable to fully measure follows across all treatments. As such, we utilized ad click-throughs as a proxy outcome measure.

Among the four ad treatments, we found that the ad with a persuasive message to follow a well-known health institution generated significantly higher click-through rates than all other conditions (Bonferroni-corrected pairwise tests, all p<0.001). Given the overall low click-through rate and high cost of an X ad-based intervention---\$978.48 for 28,582 impressions, \$3.31 per click-through, and \$123.10 per follow---we conclude that our proposed intervention is not a cost effective means to change information environments.

We argue that reporting failed interventions is essential for knowledge-building in misinformation research and preventing publication bias. Our follow nudges study suggests the following insights for future research: (1) users are unlikely to diversify their information networks when asked via ad-based interventions and (2) the infrastructure required for ad-based intervention can quickly become cost prohibitive. To compliment our experiment findings, we also ran a follow-up survey study (N=400) with X (formerly Twitter) users who express mistrust in vaccines, to understand potential reasons why the experiment participants did not click the ad treatments. Finally, our study illustrates several challenges facing future computational social science research: decreasing ability to conduct large-scale experiments under growing platform data restrictions and the unreliability of digital advertising infrastructure to aid in these experiments---particularly on X, as their Ads Manager was prone to breaking and lacked documentation.

\section{Related Work}
A large body of scholarly work proposes various misinformation interventions. Here we sort this work by individual-level interventions and environment-level interventions. Our work sits at the environment-level, as we seek to alter social network structures with the introduction of high-quality network ties.

\subsection{Individual-level interventions} Misinformation exposure and spread can occur due to individual factors. Selective exposure occurs when people seek out (mis)information which aligns with their ideology~\cite{robertson_users_2023}. Prior work in the U.S. context found that conservatives and liberals online are more likely to visit low-quality websites, if those websites were consistent with their political ideology~\cite{guess_selective_2018}. Lack of cognitive reflection, or thinking analytically about the content one encounters online, is another individual factor. Prior social media research has found that individuals with low cognitive reflection are more likely to believe and share low-quality information on online~\cite{pennycook_fighting_2019, mosleh_cognitive_2021}.

To counteract misinformation exposure and spread arising from individual factors, research has proposed interventions which prompt users to think analytically and consider the credibility of information sources. One approach builds on the design concept of nudges: "any aspect of the choice architecture that alters people’s behavior in a predictable way, without forbidding any options or significantly changing their economic incentives"~\cite{thaler_nudge_2008}. Accuracy nudges encourage users to think about the accuracy of what they are sharing. Several studies have found this intervention to reduce the sharing of misinformation, though its persistence remains unclear~\cite{pennycook_accuracy_2022, pennycook_fighting_2019, pennycook_fighting_2020}. Media literacy and pre-bunking interventions seek to educate users how to evaluate sources and be wary of circulating false claims, though pre-bunking efforts have shown limited persistence in several studies~\cite{van_bavel_using_2020, guess_digital_2020, basol_towards_2021}. 

\subsection{Environment-level interventions} Environment-level factors are also involved in misinformation exposure and spread, including user-selected defaults and social media algorithms. Online defaults occur when users configure their online environments in some way, such as configuring a web browser to open to a preferred news site~\cite{flaxman_filter_2016}. On social media, the accounts which a user chooses to follow is another type of default, as the followers' content will populate the user's feed by default. In addition to user-selected faults, social media algorithms influence what information a user sees. Online users can find themselves in filter bubbles and echo chambers, where they are surrounded by similar content from similar accounts. 

Whether or not platform algorithms directly create such conditions is contested in the literature~\cite{cinelli_echo_2021, guess_how_2023, budak_misunderstanding_2024, guess_reshares_2023}. Audit studies have found that users who consume misinformation then have more misinformation recommended to them~\cite{hussein_measuring_2020,papadamou_it_2022,tang_down_2021}. Other studies, however, have found that demand-side factors (i.e. users seeking out problematic content) outweigh algorithmic factors~\cite{guess_how_2023, budak_misunderstanding_2024}. 

Nonetheless, echo chambers nonetheless persist online and can sever users from diverse perspectives. Previous online field experiments have sought to counteract echo chambers by diversifying users' social media feeds with counter-attitudinal content~\cite{bail_exposure_2018}. In our online field experiment, we seek to counteract echo chambers by introducing new social ties to users' social networks.

To counteract misinformation exposure and spread arising from environment-level factors, social media platforms have taken various moderation actions: enacting bans on content or individuals, adding content labels, or depromoting content. After the 2016 U.S. presidential election, Facebook (now Meta) banned fake news websites from showing ads, and in the following two years, the ratio of Facebook engagements to Twitter shares of fake news decreased by 60\%---possibly due to these platform policy changes~\cite{wingfield_google_2016, allcott_trends_2019}. In addition to banning, platforms can also label problematic accounts and content, but its efficacy is mixed: some studies suggest labeling works~\cite{yaqub_effects_2020, sharevski_misinformation_2022, clayton_real_2020}, while other suggest it does not~\cite{gao_label_2018, sanderson_twitter_2021, pennycook_implied_2020}. Less is known about content depromotion, as social media companies have kept such details about their algorithms closely held. Several investigative efforts and leaked documents suggest that depromotion does reduce user engagement with problematic material~\cite{gillespie_not_2022, cameron_heres_2022}.

As social media companies have taken action to counter misinformation, however, political backlash has ensued, with U.S. conservatives alleging that social media companies are silencing right-leaning voices and U.S. liberals accusing the same companies of not doing enough to tackle misinformation~\cite{romm_republicans_2018,zakrzewski_election_2020}. Our proposed intervention of \textit{follow nudges} offers an alternative. It is not a top-down environment decision taken by technology companies, but rather a bottom-up environment decision taken by users---by prompting them to consider adding a new tie to their information network. 

\begin{figure*}[h]
\centering
\includegraphics[width=0.65\textwidth,height=0.7\textheight,keepaspectratio]{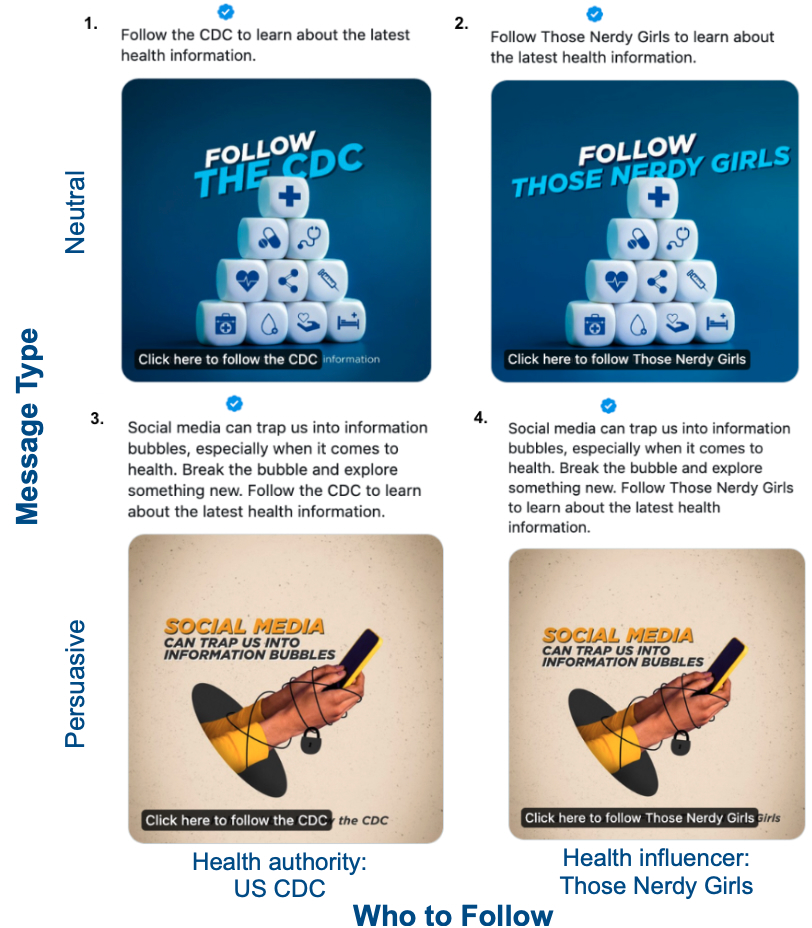}
    \captionsetup{width=0.75\textwidth}
    \caption{The four ad treatments, varied along two dimensions: (1) a neutral versus a persuasive message and (2) a request to follow a U.S. health institution versus a request to follow a health influencer}
    \label{fig:ad_creative}
\end{figure*}

\section{Experiment Design}
We conducted a between-subjects two-factor experiment utilizing the X (formerly Twitter) Ads Manager to test the effectiveness of our proposed intervention \textit{follow nudges} to encourage the following of high-quality health information sources. The participants were X users who were following health misinformation spreaders. The stimuli varied along two factors: (a) the message was either neutrally worded or employed a morally-reframed argument appealing to independence and (b) the message asked the user to either follow a health institution or to follow a health influencer, yielding four conditions. We measured the click-through rates (CTR) of ads, as well as new follows for one of the promoted health accounts.\footnote{Given major changes to the X API during this study, we were unable to fully collect followership data. We attempted several workarounds, and ultimately asked the two health accounts to share their followership lists. Only the health influencer agreed to share this information.} This experiment was approved by our university's Institutional Review Board (IRB).

\section{Participants}\label{sec:participants}
To identify followers of health misinformation spreaders, we first had to identify the health misinformation spreaders themselves. We took the following steps to identify spreaders of health misinformation on X (formerly Twitter).

\subsubsection{Identifying health misinformation stories} We collected all PolitiFact rumors that (1) originated from a blog or a tweet and (2) occurred between June 2021 and December 2023. After scraping all rumors, we filtered for rumors that were about health misinformation. PolitiFact tags rumors with topics, and we considered a rumor to be health-related if it contained one of the following tags: `abortion', `autism', `coronavirus', `drugs', `disability', `health-care', `health-check', `public-health'. We then manually verified that the scraped originating URL of the rumor was indeed health misinformation, as opposed a fact-check debunking a health misinformation claim. If the originating URL was a tweet, we verified that the tweet had not since been deleted. We then filtered to following PolitiFact-assigned truth values: `pants on fire', `false', and `mostly false.' This filtering process yielded 89 URLs which directed to tweets or blog posts containing health misinformation. 

\subsubsection{Identify whether spreaders of the health misinformation stories have X (Twitter) accounts}
Among the identified health misinformation spreaders, we then identified the `speaker' or source of the misinformation story, and then looked to see if they had an X account. We filtered out any suspended accounts and elected representatives. This resulted in 38 eligible spreaders, who had shared 52 of the 89 misinformation stories identified in the previous step.

\subsubsection{Identify prolific health misinformation spreaders with large followings} To identify X (formerly Twitter) accounts with large audiences and that frequently shared health misinformation stories, we created an index which considered each account's metadata and posting history. The index equally weighed the number of followers and the number of health misinformation stories shared: 0.5*rank(followers) + 0.5*rank(number of health misinformation stories shared). We chose the top five health misinformation spreaders who scored the highest on this index: Jack Posobiec (@JackPosobiec), Candance Owens (@RealCandaceO), Charlie Kirk (@CharlieKirk11), The Gateway Pundit (@gatewaypundit), and Steve Kirsch (@stkirsch).
 
\subsubsection{Identify followers of spreaders of health misinformation stories on X (Twitter)}
For each of the five spreaders, we pulled a maximum of 450,066 followers, i.e., the lowest number of followers of the five spreaders at the time the data was collected. After pulling 2,250,330 followers across the 5 spreaders, we de-duplicated and randomly sampled four groups of 20,000 for our treatment groups.\footnote{Our power analysis required 6,280 participants per ad. When constructing the treatment groups in the X Ads Manager, we uploaded 20,000 users per audience list to account for the X Ads Manager's processing losses: typically only 8-9,000 of 20,000 uploaded users were successfully found. This conservative approach ensured we would meet our 6,280 target after processing. X did not provide documentation on what accounted for this processing loss.} Finally, when constructing the treatment groups via the Custom Audience Lists in the X Ads Manager, we filtered to only accounts geolocated to the United States.

\section{Stimuli}\label{sec:stimuli} We varied our \textit{follow nudges} along two factors: the message and the account we ask the user to follow. This resulted in four treatment conditions: (1) an ad to follow a U.S. health institution using a neutral message, (2) an ad to follow a U.S. health institution using a persuasive message, (3) an ad to follow a health influencer using a neutral message, and (4) an ad to follow a health influencer using a persuasive message. We worked with a graphic designer to develop illustrations for the ads. We decided to vary the ad graphic with the message dimension, resulting in two ad illustrations: one aligned with the neutral message and another aligned with the persuasive message. See Figure~\ref{fig:ad_creative} for the resulting ad creatives. 

\subsubsection{Factor 1: Message Type} To vary the message, we have a neutral message (i.e., a simple message to follow a new account) as well as a persuasive message. The neutral message reads: ``Follow the (health institution or health influencer) to learn about the latest health information.''

To craft the persuasive message, we drew upon existing work in public health messaging on values-aligned messages~\cite{shen_persuading_2023}. Values reframing (or moral reframing) refers to crafting a message to align with the values of the target audience.~\citet{kaplan_moral_2023} found that reframing mask-wearing as an expression of loyalty to one's country decreased anti-mask sentiment.~\citet{bokemper_testing_2022} reframed refusal to social distance as a reckless act rather than a brave one, and found that this increased participants' intention to social distance. 

In our study, we reframed following a health authority as an independent act to `break out' of a user's existing social media information bubble. The persuasive, values-reframed message reads: ``Social media can trap us into information bubbles, especially when it comes to health. Break the bubble and explore something new. Follow the (health institution or health influencer) to learn about the latest health information.'' We chose this wording in part as it is reminiscent of the phrase ``do you own research'' which has been utilized by online groups to encourage distrust of health institutions and other authoritative sources. Usage of this phrase as been found to be associated with belief in COVID-19 misinformation~\cite{chinn_support_2023}. By suggesting to online users to `break the bubble and explore something new' we are in part capitalizing on online users' drive to `do one's own research' in the internet age, but encouraging them to look to reliable sources as a place in that process.

\subsubsection{Factor 2: Account to Follow} To vary the account to follow, we selected a health institution as well as a health influencer. Given declining trust in government and media institutions~\cite{ognyanova_misinformation_2020}, we sought a non-institutional health messenger. For a U.S. health institution, we selected the Centers for Disease Control and Prevention (CDC)---the national public health agency responsible for monitoring disease outbreaks and other health threats, as well as developing evidence-based public health messaging. For a U.S. health influencer, we selected Those Nerdy Girls, an all-woman team of scientists who translate emerging public health literature for a lay audience~\cite{leininger_fight_2022}. Those Nerdy Girls formed in March 2020, at the start of the COVID-19 pandemic, and continue to create public health messaging on current issues, such as the ongoing U.S. measles outbreak as of the writing of this paper~\cite{markin_whats_2025}.


\subsection{Procedure}
To deliver the \textit{follow nudges} to followers of health misinformation spreaders, we utilized targeted digital advertisements via the X Ads Manager. For each of the four ad treatments, we ran an ad campaign with a Custom Audience List,\footnote{\url{https://business.x.com/en/help/campaign-setup/campaign-targeting/custom-audiences/lists}} which allowed us to display each ad to the respective treatment groups.

Upon clicking the ad treatments (1) or (3), the participant was navigated to the account of the CDC (@CDCgov). Upon clicking the ad treatments (2) or (4), the participant was navigated to the account of Those Nerdy Girls (@DearPandemic). Specifically, we used the Follow Button Web Intent URLs.\footnote{\url{https://developer.x.com/en/docs/x-for-websites/follow-button/guides/web-intent-follow-button}} This link directs to the account's page with a prompt box asking, ``Do you want to follow @Account?'' and a follow button directly below. 

We launched all four ad campaigns on the same day and ran the campaigns until we met the sample size as determined in our power analysis---6,280 participants per treatment group. The four ad campaigns ran between February 5 and February 7, 2025.\footnote{The experiment was initially planned to run in Fall 2024. Due to technical issues with the X Ads Manager, the experiment was delayed until February 2025---shortly before a leadership transition and significant organizational changes at the CDC. We discuss this further in Section 7.} We tracked the number of click-throughs for each ad using the URL shortener Linkly.\footnote{\url{https://linklyhq.com/}}  We primarily rely on this proxy outcome measure for our analysis. The ground truth outcome measure--following the promoted account--was ultimately made impossible by the X API, which significantly degraded the functionality of the previous Twitter API. We provide further discussion of the limited X API and its implications for future work in Section 7. In lieu of measuring new follows ourselves, we contacted both the CDC and Those Nerdy Girls and asked if they would share their follower lists pre- and post-experiment. The CDC declined to, while Those Nerdy Girls agreed. As such, we were able to track actual follows for the health influencer, but not the U.S. health institution.

\section{Findings}
\begin{figure}[h]
\includegraphics[width=0.5\textwidth,height=\textheight,keepaspectratio]{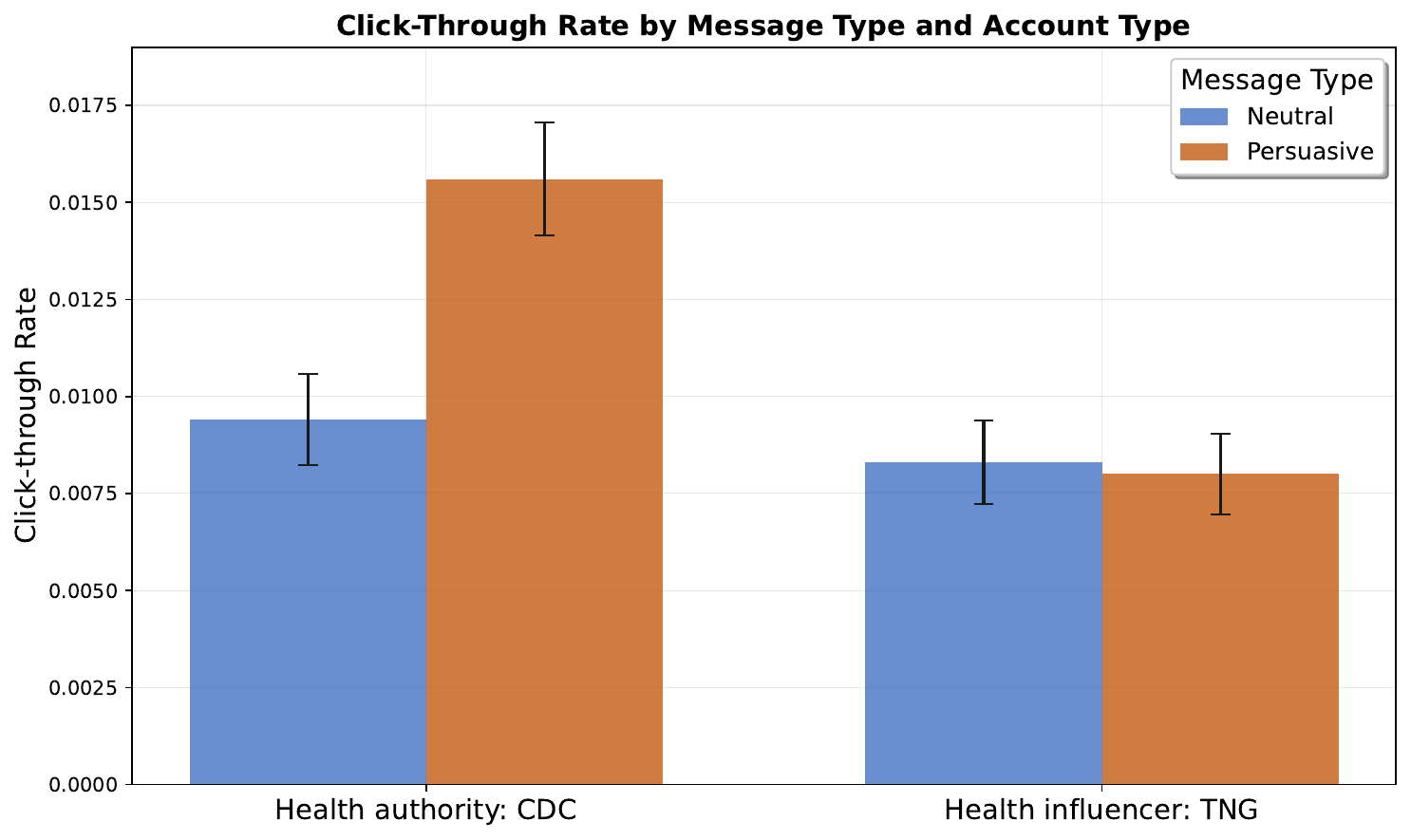}
    \captionsetup{width=0.5\textwidth}
    \caption{Comparison of click-through rates across ad treatments with standard error bars}
    \label{fig:ad_results}
\end{figure}

Below we discuss the success of each ad treatment in terms of click-through rates (CTR) and actual follows. See Table~\ref{tab:all_results} for a summary of exposures, clicks, and follows across experiment arms and Figure~\ref{fig:ad_results} for a visual comparison of the click-through rates. 

\begin{table}[]
\begin{tabular}{l|rrrr}
                         & \multicolumn{1}{c}{\textbf{Ad 1}} & \multicolumn{1}{c}{\textbf{Ad 2}} & \multicolumn{1}{c}{\textbf{Ad 3}} & \multicolumn{1}{c}{\textbf{Ad 4}} \\ \hline
\textbf{Message Type}    & Neut                              & Neut                              & Pers                              & Pers                              \\
\textbf{Account}         & CDC                               & TNG                               & CDC                               & TNG                               \\
\textbf{Exposures}       & 6806                              & 7075                              & 7305                              & 7396                              \\
\textbf{Clicks}          & 64                                & 59                                & 114                               & 59                                \\
\textbf{CTR}             & 0.94\%                            & 0.83\%                            & 1.56\%                            & 0.80\%                            \\
\textbf{Cost per click}  & \$3.74                            & \$4.14                            & \$2.17                            & \$4.19                            \\
\textbf{Follows$^\dagger$}         & --                                & 2                                 & --                                & 1                                 \\
\textbf{Follow rate$^\dagger$}     & --                                & 0.03\%                            & --                                & 0.01\%                            \\
\textbf{Cost per follow$^\dagger$} & --                                & \$122.23                          & --                                & \$247.07                         
\end{tabular}
\caption{Summary of experiment results by condition}
$^\dagger$ Complete follow data unavailable due to X API restrictions
\label{tab:all_results}
\end{table}

\begin{figure*}[h]
\centering
\begin{subfigure}{0.48\textwidth}
    \centering
    \includegraphics[width=\textwidth,height=0.8\textheight,keepaspectratio]{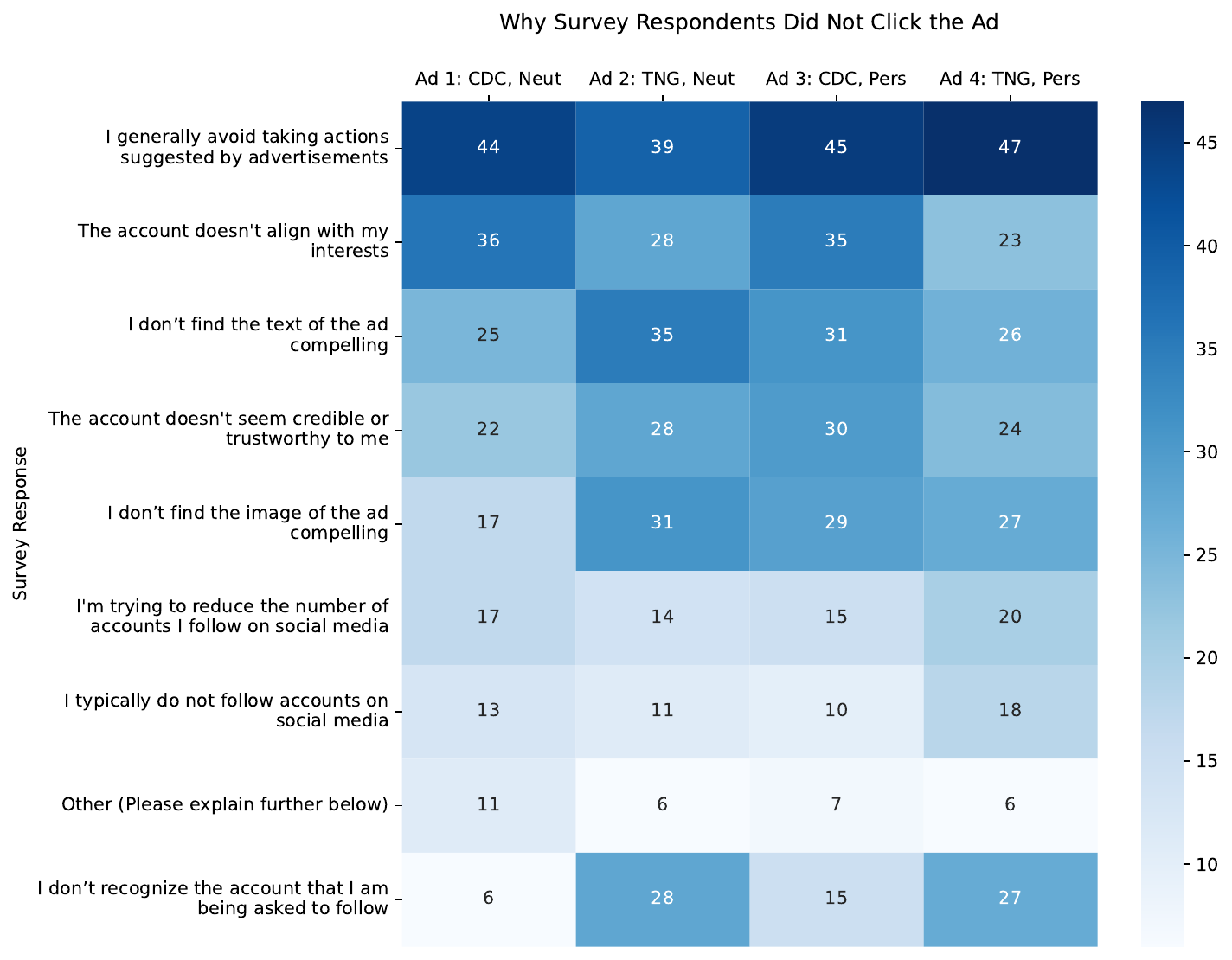}
    \label{fig:heatmap_results_a}
\end{subfigure}
\hfill
\begin{subfigure}{0.48\textwidth}
    \centering
    \includegraphics[width=\textwidth,height=0.8\textheight,keepaspectratio]{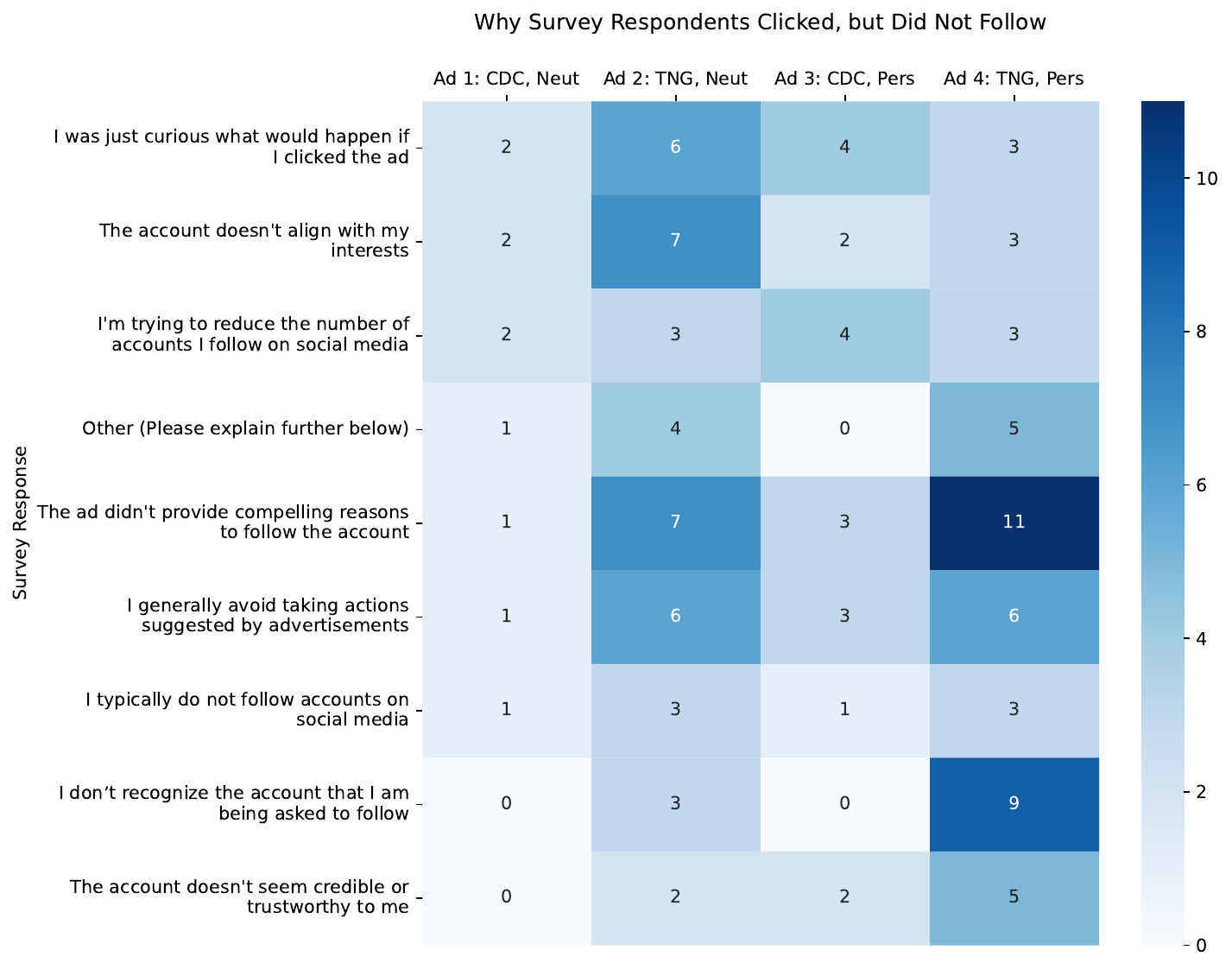}
    \label{fig:heatmap_results_b}
\end{subfigure}
\captionsetup{width=0.85\linewidth}
\caption{Survey results for two subsets of participants. On the left, a heatmap of responses from those who did not click the ad. On the right, a heatmap of responses from those who clicked the ad, but chose not to follow the promoted account.}
\label{fig:heatmap_comparison}
\end{figure*}

\subsection{Click-through Rates}The four ads received 28,582 impressions and 296 click-throughs, resulting in an overall click-through rate (CTR) of 1.04\%. This CTR is inline with industry reports on average click-through rates for X (formerly Twitter) ad campaigns, which have been reported to range from 0.80\% to 1\%~\cite{gardner_social_2023}. The CTR of 1.04\% represents an upper bound of potential follows, since users must click the ad before they can follow an account. Major changes to the X API during our study prevented us from directly measuring follows across all treatments, requiring us to rely upon click-through rates as a proxy outcome measure. We conclude that our intervention is ineffective: since clicks provide only an upper bound on actual follows, the true follow rate likely even lower than 1.04\%.

To assess the highest performing ad treatment, we conducted an omnibus chi-square test followed by post-hoc pairwise comparisons. The omnibus test confirmed significant differences across all four conditions. Pairwise two-proportion z-tests with Bonferroni correction ($\alpha$ = 0.05/6 = 0.0083) revealed that the Ad Campaign 3 (a persuasive message to follow the CDC) significantly outperformed all other conditions: Ad Campaign 1 (a neutral message to follow the CDC, CTR=0.94\%, p<0.001), Ad Campaign 2 (a neutral message to follow Those Nerdy Girls, CTR=0.83\%, p<0.001), and Ad Campaign 4 (a persuasive message to follow Those Nerdy Girls, CTR=0.80\%, p<0.001). See Table \ref{tab:all_results} for additional details.

\subsection{Following rates}
In terms of tracking actual follows, we were unable to collect this data via the new X API---something that would have been a trivial task with the former Twitter API. We discuss various attempted workarounds in Section 7. We ultimately contacted the CDC and Those Nerdy Girls directly and asked if they would share their follower lists pre- and post-experiment. The CDC declined, while Those Nerdy Girls agreed to share their follower lists. With the follower lists downloaded before and after the experiment, we found that Those Nerdy Girls received only three new follows following our experiment: two follows from the neutral message condition and one follow from the persuasive message condition. This yielded conversion rates of 0.03\% for the Ad Campaign 2 (a neutral message to follow Those Nerdy Girls) and 0.01\% for Ad Campaign 4 (a persuasive message to follow Those Nerdy Girls). See Table \ref{tab:all_results} for additional details. These low conversion rates, and our inability to measure follows for the CDC campaigns, prevented the detection of statistical differences across experimental conditions. In the next section, we examine the cost-effectiveness of the \textit{follow nudge} interventions. 

\subsection{Cost analysis}
Running targeted digital ads on social media comes with non-trivial expense. Table~\ref{tab:all_results} summarizes the costs across all four ad campaigns. Ad Campaign 1 (a neutral message to follow the CDC) cost \$239.33, yielding \$0.0351 per impression and \$3.74 per click-through. Ad Campaign 2 (a neutral message to follow Those Nerdy Girls) cost \$244.46, yielding \$0.0346 per impression and \$4.14 per click-through. Ad Campaign 3 (a persuasive message to follow the CDC) cost \$247.63, yielding \$0.0339 per impression and \$2.17 per click-through. Ad Campaign 4 (a persuasive message to follow Those Nerdy Girls) cost \$247.07, yielding \$0.0334 per impression and \$4.19 per click-through. 

The cost per actual follow was substantially high. Ad Campaign 2 resulted in two follows for Those Nerdy Girls, at \$122.23 per follow. Ad Campaign 4 resulted in one follow, at \$247.07 per follow. These costs demonstrate the intervention's lack of cost-effectiveness. In total, all four ad campaigns cost \$978.48 for 28,582 total impressions garnered over 3 days---\$0.0342 per impression across campaigns and \$163.84 per actual follow of a high-quality health source.

\subsection{Follow-up survey}
To better understand why participants exposed to the \textit{follow nudge} ads chose not to click on them, we conducted a follow-up survey, which was also approved by our university's Institutional Review Board (IRB). The survey was fielded using Qualtrics and Prolific with 400 respondents who reported using X (formerly Twitter) at least once per month and who had expressed some degree of vaccine skepticism or mistrust in science. We utilized both Prolific's prescreened audiences and in-study screening feature, which allows for non-qualifying participants to be screened out while still receiving a small payment.\footnote{\url{https://www.prolific.com/resources/what-s-new-expanded-quotas-in-study-screening-and-smarter-quality-controls}} This ensured to the best of our ability that our survey participants approximated a similar population to our experiment participants, i.e. those who are reachable through social media and likely to hold reservations about vaccination. See Appendix~\ref{sec:survey_prescreening} for screening questions.

Survey participants were presented an ad and asked to consider it for 10 seconds. After, they decided whether or not they would click on the ad. For those who did not click, they were asked to explain their decision through multiple choice and open-ended questions. For those who clicked, but did not follow the promoted account, they were also asked to explain their decision through multiple choice and open-ended questions.

We received roughly 100 responses across each of the four ad treatments, using the Qualtrics randomizer feature.\footnote{Ad 1 received 99 survey responses, Ad 2 received 99 survey responses, Ad 3 received 100 survey responses, and Ad 4 received 102 survey responses.} Survey participants were compensated at a rate of \$15 per hour. The median response time was 2 minutes and 32 seconds.

Figure~\ref{fig:heatmap_comparison} present results for our two subsets of survey participants: (A) those who did not click the ad and (B) those who clicked but chose not to follow the promoted account. Overall, responses did not differ greatly between these two groups. Both groups expressed a general reluctance to act on advertisements and being uncompelled by the ad creative. The heatmap shows these patterns were generally consistent across the four ad treatments.

There are two notable places where survey responses diverged: (1) Participants noted that the CDC account does not align with their interests more so than the health influencer account, Those Nerdy Girls (TNG). (2) Participants seemed unwilling to click on the health influencer ads given that they were unfamiliar with the account. These results suggest that the credibility and familiarity of the promoted health accounts impacted participants' responses: the established health institution was penalized for its notoriety, while the newer health influencer account was penalized for its obscurity. The open-ended responses corroborated this trend, with many participants expressing mistrust and even vitriol towards the CDC, while voicing ambivalence towards TNG.

\section{Discussion}
Considering the high cost and low conversion rates observed across all conditions, this study provides empirical evidence that ad-based \textit{follow nudges} are not cost-effective for improving online information environments affected by health misinformation at scale. We discuss several implications below. 

\subsection{A high-powered field experiment suggests misinformation networks are resistant to nudges encouraging new social ties.} We conducted a large-scale online field experiment on X (formerly Twitter) with real followers of real health misinformation spreaders, using the X Ads Manager. This provides a strong ecological validity to our finding that ad-based \textit{follow nudges} are largely ineffective in convincing followers of misinformation to follow spreaders of high-quality health information. Beyond ad-based nudges, there could be other interventions that prove more effective. Our current work, however, suggests that misinformation networks may be entrenched in the following ways: online users who follow spreaders of health misinformation are \textit{unlikely to diversify their information networks via ad-based nudges}, and the digital advertising infrastructure needed quickly becomes cost-prohibitive---preventing this intervention from being viable at scale. This resistance is compounded by the fact that followers are \textit{unlikely to unfollow those they have already followed, even spreaders of health misinformation}~\cite{kivran-swaine_impact_2011, ashkinaze_dynamics_2024} . In sum, followers in health misinformation networks appear resistant to follow new high-quality sources and also reluctant to unfollow existing low-quality sources. 

\subsection{The last large-scale network experiment on Twitter?} Following the social media giant's ownership change, we faced various roadblocks and hurdles to conducting this study. These platform changes will likely obstruct future computational social science research on X. 

\subsubsection{Tracking the DV} The first challenge was the X API.\footnote{https://docs.x.com/x-api/introduction} Compared to the previous Twitter API, the X API does not provide access to accounts' followers at the Free, Basic, or Pro tiers. The Enterprise tier---which starts at \$42,000 a month---purports to have a `Follows Lookup.'~\footnote{As our research team does not have Enterprise API access, we are unable to confirm whether followership is available at this tier.
\url{https://docs.x.com/x-api/users/follows/introduction}} With the Basic API (\$200 a month), we could not access follower lists for the CDC and Those Nerdy Girls. This would have a simple---and free---API call under the old Twitter API regime. We attempted to build a workaround that involved multiple bot accounts and taking daily screenshots of the ``Followers You Know'' page---which ultimately proved unfeasible. See Appendix~\ref{sec:x_api} for more details. As a result, we had to rely on a proxy outcome measure: click-throughs on the ad, rather than actual follows of the health account. Since users must click on the account to follow it, measuring the ad click-throughs captured an upper bound on the true follow rate, which was likely lower.

\subsubsection{Identifying participants} Participant identification occurred prior to the API downgrade and involved a process that is not longer possible under the new API regime. In addition, we had planned to filter to active accounts immediately prior to running the experiment. See Appendix~\ref{sec:account_filtering} for details. With the downgraded API, we were not able to conduct these checks. Our only `active account' check came from X Ads Manager, which purports to filter out inactive accounts in the Custom Audience List feature.

\subsubsection{Running an ad campaign} X's Ads Manager, the platform's digital advertising infrastructure, presented three main difficulties: (1) unclear limitations on which accounts can run ad campaigns, (2) incomplete data on ad impressions, and (3) a tendency to break without explanation. First, the approval process to run advanced ad campaign is opaque. We had an established Twitter (pre-X) account with blue check verification, which passed the ads `premium verification.' We had several other less established accounts, and these accounts were unable to run advanced ads campaigns. Second, X does not provide the metric for unique user impressions. We confirmed with X Ads Support several times that their ad impressions metric represents total views, not unique users. An ad may receive 100 impressions, but it may only have been seen by 80 people. Third, to run ads to four separate treatment groups, we relied on Custom Audiences, which enables an ad campaign to be shown to a predefined list of X usernames.\footnote{\url{https://business.x.com/en/help/campaign-setup/campaign-targeting/custom-audiences}} This functionality was broken from October 2024 until January 2025, delaying the project considerably. Despite the first author providing extensive evidence of the problem (including screenshots and screen recordings), X Ads Support could not explain why the tool was broken and did not provide notification when the tool began working again in January 2025.

\subsection{Limitations \& Future Work}
Our study contains several limitations. First, our collection of health misinformation spreaders and their followers is skewed to the English-language speaking North American context, given that our groundtruth for health information fact checks was PolitiFact, which is focused primarily on North America. 

Second, we tested our ad-based social network misinformation intervention only on one social media platform (X), during a time when the platform was undergoing various changes to how content was recommended and moderated~\cite{conger_how_2023}. Future work could explore whether ad-based social network interventions show potential other social media networks using field experiments similar to this one, or in lab environments utilizing synthetic social networks. Future work must also consider the long-term utility of network-based interventions, as social media companies continue to shift from social network-based content recommendations, to algorithmic-based recommendations which are less tethered to who follows who~\cite{kantrowitz_surprising_2023}.

Third, the selection of the CDC for the health authority account would have been a relatively straightforward choice in previous years. Our experiment was planned for Fall 2024, but was delayed due to the X Ads Manager technical issues discussed above. We ultimately ran the experiment in February 2025---just before leadership transitions at HHS and CDC with the incoming administration. In the ensuing months, the CDC experienced heightened media attention, as it underwent personnel layoffs, budget cuts, policy reversals on vaccine recommendations, and an active shooter event at its headquarters in Atlanta, Georgia~\cite{kekatos_mass_2025}. 

The heightened awareness of the CDC in the public consciousness might have contributed to some of Ad Campaign 3's success in garnering click-throughs. From the qualitative survey results, however, it appears there is an undercurrent of public distrust and dislike for the CDC. Future work might explore what other high-quality health information sources might be good candidates for this social network-based intervention. Rather than asking followers of health misinformation to follow a health institution experiencing heightened public scrutiny, the \textit{follow nudges} could perhaps nudge users towards lesser-known health institutions.

\section{Conclusion}
We conducted a large-scale field experiment on X (formerly Twitter) with users who follow accounts that spread health misinformation. We tested four ads treatments, varied on two dimensions: a neutral message versus a persuasive message appealing to values of independence, and a request to follow a U.S. health institution (the CDC) versus a request to follow a health influencer (Those Nerdy Girls). The ad with a persuasive message to follow the well-known health institution generated significantly higher click-through rates than all other conditions. Given, however, overall low conversion rates across treatments and the high cost of utilizing digital advertising infrastructure, we conclude that our proposed intervention of follow nudges---at least in its current ad-based format---is not a cost-effective means to improve information environments at scale.

\section{Ethics Statement} There are ethical implications to running an online experiment with followers of health misinformation spreaders. We took several steps to mitigate risks. First, we refrain from sharing any identifiable information about these followers. Second, our experiment design is minimally invasive, as it is conducted via digital advertising campaigns on social media---a treatment which the participants can easily choose to ignore. Third, in our follow-up survey, we collected no identifiable survey participant information. Finally, as discussed in the Related Works section, one advantage of our proposed intervention \textit{follow nudges} is its bottom-up (as opposed to top-down) nature. Users are encouraged to change their online information environments by adding a new social tie. Other misinformation interventions, such as banning users or content, are more top-down in nature, and may be considered more invasive approaches to counteracting problematic content online.

\begin{acks}
[Anonymized for review]
\end{acks}

\bibliographystyle{ACM-Reference-Format}
\bibliography{references}

\appendix
\section{Appendices}

\subsection{Survey prescreening questions}
\label{sec:survey_prescreening}
We asked the following screening questions in our survey so that the survey participants would resemble our experiment participants. Participants who used X once a month or more, and expressed some distrust in either scientists to act in the public interest or distrust in scheduled vaccines for children passed the screening process.

\begin{itemize}
    \item \textbf{Screening question 1:} How often do you use the social media platform X (formerly Twitter)?
    \begin{itemize}
        \item Daily
        \item Once a week
        \item Once a month
        \item A few times a year
        \item Rarely
        \item I do not have an account
    \end{itemize}
    
    \item \textbf{Screening question 2:} On a scale of 1 to 5, how much confidence do you have in scientists to act in the best interests of the public?
    \begin{enumerate}
        \item None
        \item Not too much
        \item Unsure
        \item A fair amount
        \item A great deal
    \end{enumerate}
    
    \item \textbf{Screening question 3:} On a scale of 1 to 7, please rate to what extent you agree with the following statement: ``Scheduled vaccines for children are generally safe.''
    \begin{enumerate}
        \item Strongly disagree
        \item Disagree
        \item Somewhat disagree
        \item Neither agree nor disagree
        \item Somewhat agree
        \item Agree
        \item Strongly agree
    \end{enumerate}
\end{itemize}




\subsection{Attempted X API workaround}
\label{sec:x_api}
Given the severely degraded X API that did not provide followership data, we attempted a workaround which would exploit the Followers You Know page. If we followed all the study participants, and then semi-regularly visited the Followers You Know page for both the CDC and Those Nerdy Girls, then we might have a chance at accurately capturing the DV for this study. This workaround would require three steps: (1) rehydrating the followers so that we found their username from the X user id, (2) following all of the followers/participants (n=20,000), and then (3) check the CDC and TNG accounts multiple times a day to see if new account appeared in the Followers You Know page. We ran into several issues at all three steps. At Step 1, we faced slower rehydration rates then previous Twitter API allowed. At Step 2, we encountered limits on the amount of accounts we could follow: 400 accounts per day, and then no more following after 5,000 until others started following you back. We had multiple X accounts frozen. At Step 3, the ``Followers You Know`` page does not allow you to scroll all the way to the bottom if you have more than 20 accounts in common. As such, we developed a script to take a screenshot of the Followers You Know page multiple times a day and parse out the new accounts at top. Ultimately, our ``Followers you know`` workaround was not viable, given our required sample size (20,000) and X's various limitations. Again, collected this data would have a been simple---and free---API call under the old Twitter API regime. 

\subsection{Planned Account Activity Filtering}
\label{sec:account_filtering}
Prior to the X API change, we had planned to a final filtering of misinformation follower accounts just before the ad experiment, to ensure that these accounts were active. The filtering mechanisms were: (1) Last tweet was in the last 14 days or more recent; (2) Ideology could be estimated via~\cite{barbera_birds_2015}; (3) Follower and following count was greater than or equal to 20, (4) tweet count was greater than or equal to 10. This filtering step was made impossible by the API downgrade.
\end{document}